\documentstyle[12pt]{article}

\newcommand{\veps}{\varepsilon}
\newcommand{\vphi}{\varphi}
\newcommand{\D}{\partial}
\newcommand{\g}{{\rm g}}
\newcommand{\fvec}{{\bf f}}
\newcommand{\avec}{{\bf \alpha}}
\newcommand{\tvec}{{\bf t}}
\newcommand{\tauvec}{{\bf t}}
\renewcommand{\tau}{t}
\newcommand{\text}{\,}

\newcommand{\Integer}{{\bf Z}}

\newcommand{\Real}{{\bf R}}

\renewcommand{\th}{{\rm th\,}}

\newcommand{\intl}{\int\limits}
\newcommand{\suml}{\sum\limits}
\newcommand{\prodl}{\prod\limits}
\newcommand{\be}{\begin{equation}}
\newcommand{\ee}{\end{equation}}
\newcommand{\ba}{\begin{eqnarray}}
\newcommand{\ea}{\end{eqnarray}}


\begin{document}

\vskip2cm

\title{\Large\bf A New Class of Elliptic Finite-Gap\\
                 Densities of the Polar Operator\\
                 and Stationary Solutions of the Harry Dym equation}
\vskip1cm
\author{L.~A.~Dmitrieva, D.~A.~Pyatkin\\
         \it Department of Mathematical\\
         \it \& Computational Physics, St.-Petersburg State University,\\
         \it 198504 St.-Petersburg, Russia\\
             e-mail: mila@JK1454.spb.edu\\
                     danila@DP2627.spb.edu}
\date{}

\maketitle

\abstract{
A new family of one- and two-gap elliptic densities of the polar
operator has been constructed by modifying the so-called "higher
time approach" to constructing finite-gap solutions of the
Harry Dym equation. Auto-B\"{a}cklund transformations of the
obtained stationary solutions have been constructed and their
properties have been studied.}

\section{Introduction}

The present work is devoted to constructing elliptic one- and
two-gap densities of
the polar operator
\be
   L=-r^2(x)\,\frac{d^2}{dx^2}, \label{polar}
\ee
with periodic function $r(x)$: $r(x+T)=r(x)$ on the real axis
$x\in\Real$. Function
$r(x)$ is referred to as density of the polar operator.
It is easy to see that the operator $L$ is intimately connected
with the string equation
\be
   \psi''_{xx}+\lambda\,\rho(x)\,\psi=0 \label{problem}
\ee
provided that $\rho(x)=r^{-2}(x)$ where $\rho(x)$ is the density of the string and
$\psi$ the amplitude of it's oscillation. We construct and study functional properties
of some finite-gap densities of the operator (\ref{polar}), namely 1-gap and 2-gap
elliptic densities%
    \footnote{${}^)$ we call a density elliptic if it can be expressed in terms of
                     elliptic functions.})
of the operator $L$.

Despite classical character of the spectral problem (\ref{problem}) for the
operator $L$ and a great number of papers devoted to the string equation
(see works \cite{KR}, \cite{KRA}, \cite{ATK}, \cite{KUP}, \cite{DK} and reference
therein to name a just few),
hitherto there weren't any explicit analytical formulae for the finite-gap
densities $r(x)$ with the exception of expression for smooth 1-gap density
constructed explicitly in \cite{DM1}. Here by explicit representation we assume
representation of the form
$$
  r(x)=R(y),
$$
where $y=y(x)$ is solution of the functional equation
$$
  x=F(y),
$$
and the functions $R$ and $F$ are given explicitly.

The absence of analytical formulae for the finite-gap
densities doesn't seem strange despite the relation
between the string equation (\ref{problem}) and the
one-dimensional Shr\"odinger equation
\be
   -\Omega''_{yy}+u(y)\,\Omega=\lambda\,\Omega \label{Schredinger}
\ee
for which the class of finite-gap potentials is well known
\cite{ITS}. This relation has the form
\ba
\nonumber
&  u(y)={}^1/{}_2\,r^2_x(x)-{}^1/{}_4\,r(x)\,r_{xx}(x)=
   \left(R^{-1/2}\right)_{yy}\,R^{1/2} &\\
\nonumber
&  y=\intl^x \frac{dx'}{r(x')} &\\
\nonumber
&  \psi(x,\lambda)=\Omega(y,\lambda)\,r^{{}^1/{}_2}(x). &
\ea
So if the density $r(x)$ is known then the potential $u(y)$ can
be reconstructed easily. However the above relations can not
be used for effective constructing
the densities of the polar operator directly.

   At the same time the fact of finite-gap densities' existence was proved
by M.~G.~Krein more than 40 years ago. In the paper \cite{KR} a theorem was stated
(without proof) asserting that there exist $2^{N-1}$ $N$-gap densities
of the string equation provided that the left edge of the operator's $L$
zonal spectrum is situated at the point $\lambda=0$. The last condition
in the Krein's theorem is essential%
         \footnote{${}^)$ note that location of the Schr\"odinger zonal spectrum's edge
                          in the point $\lambda=0$ is {\bf not} essential}),
since in case of it's violation generally speaking the number of $N$-gap densities
is greater than $2^{N-1}$. In the present paper we illustrate this effect by
studying arrangement of the operator's spectrum in regard to the point
$\lambda=0$ and in case of 1- and 2-gap {\it elliptic\/} densities show that
the set of all 1- and 2-gap densities is considerably wider and contains
also some singular densities which were out of consideration in \cite{KR}.

   In order to construct smooth and singular 1- and 2-gap densities of the
operator $L$ we make use of the relation of linear spectral problem for the polar
operator $L$ with solutions of nonlinear evolutionary integrable Harry Dym
equation \cite{CAL}:
\be
   r_t+r^3\,r_{xxx}=0,\quad r=r(x,t), \label{HDeq}
\ee
and it's higher analogues.%
         \footnote{${}^)$ hierarchy of the Harry Dym equations is described later})
This relation is based on the fact that \cite{DM1} holding the time variable $t$
in the solution of the higher Harry Dym equation the resulting function turns out
to be finite-gap density of the string equation.
Because of this and the purpose of our paper (as we interested mostly in
densities of operator $L$) we will not distinguish between densities of the string
equation and solutions of Harry Dym equation.

For the first time the finite-gap
solutions of the Harry Dym hierarchy were constructed in \cite{DM1} using higher times
method developed in \cite{DM2}. In \cite{DM1} the finite-gap solutions were constructed in
terms of multi-dimensional Riemann $\Theta$-functions and needed considerable
effectivization. In the present paper we modify the scheme for construction
finite-gap solutions of the Harry Dym equation and reduce the $\Theta$-functions
form in case of 2-gap densities to elliptic functions. This reduction becomes
possible after the Riemann surface of the polar operator's Bloch functions
is chosen to possess symmetry of the special sort.

   Along with higher times technique we use auto-B\"{a}cklund transformation
\cite{RN} for higher equations of Harry Dym hierarchy.
We show that auto-B\"{a}cklund transformation allows one
to construct new 1-gap densities of the polar operator. We also construct
auto-B\"{a}cklund transformations in 2-gap case and study their properties
leaving the property of spectrum preservation as hypothesis.

\section{Modified scheme for construction of finite-gap solutions
         of Harry Dym equation.}\label{scheme}

   The scheme for construction of $N$-gap ($N<\infty$) solutions of the
Harry Dym equation proposed in \cite{DM1} is the following. The $N$-gap
solution of the eq. (\ref{HDeq}) $r(x,t_1)$ is assumed to depend on $N-1$
additional parameters --- higher times $t_2,\dots,t_N$ in such a way that
solution $r(x,\tvec)\equiv r(x,t_1,\dots,t_N)$ with respect to $t_m$,
$1\leq m\leq N$, obeys the $m$-th higher Harry Dym equation:
\be
   r_{t_m}+r^3\,(\D_x^3 r I r)^m\,r^{-3}\,r_x=0, \quad m=1,\dots,N.
   \label{hierarchy}
\ee
Here the operator $I$ is defined by equations $I\,\D_x r^{-1}=r^{-1}-1$
and $I\,\D_x\vphi(x,\tvec)=\vphi(x,\tvec)$, if
$\vphi$ is any differential polynomial of $r$.

   The $N$-gap solutions of the system (\ref{hierarchy}) were obtained
\cite{DM1} in the form:
\be
   r(x,\tvec)=R(y,\tvec)=\frac1{1+\Big<\avec,\fvec(y,\tauvec)\Big>} \label{general}
\ee
where $y=x+\veps(x,\tauvec)$.
The phase function $\veps(x,\tvec)$ is related with solution
$r(x,\tvec)$ by equation $\veps_x=1/r-1$ and finally should be determined
by solving functional equation
$$
  \veps(x,\tvec)=E(y,\tauvec)=\intl_0^y (1-R(y',\tauvec))\,dy'.
$$

   In (\ref{general}) by
$<\cdot,\cdot>$ we denote the inner product in $\Real^N$.
The constant vector $\avec$ and the function $\fvec(y,t)$ have been described
explicitly in \cite{DM1} in terms of $N$-dimensional Riemann $\Theta$-function
and the periods of holomorphic differentials on the hyperelliptic curve
\be
   \Lambda^2=\prodl_{i=0}^{2N}(\lambda-E_i),\quad E_i\neq E_j \label{curve}
\ee
with real numbers $E_i$ coinciding with edges of spectral gaps of the
polar operator (\ref{polar}) associated with Harry Dym equation.

   Remind that explicit expressions for $f_m(y,\tauvec)$ were obtained
in \cite{DM1} basing on the relation
$$
  f_m=\veps_{t_m}(x,\tvec),           
$$
and on the link \cite{RN} between Harry Dym and
Korteweg-de Vries (KdV) hierarchies. Namely let $u(y,\tau_1,\dots,\tau_N)$ be
a solution of the first $N$ equations of the KdV hierarchy, then
\be
   u_{\tau_m}={}^1/{}_2\,\D_y\,\veps_{t_{m+1}},\quad m=1,\dots,N.\label{um1}
\ee
In particular $u=-{}^1/{}_2\,\veps_{t_1}$.

  In the present paper we modify the scheme of construction of finite-gap
solutions of the Harry Dym equation using Hamiltonian structure of the
KdV hierarchy. Namely the system of the first $N$ equations of the hierarchy
$$
  u_{\tau_m}=-L^m\,u_y,\quad m=1,\dots,N,
$$
where
$$
  L=\D_y^2-4u-2u_y\intl^y dy\,\cdot
$$
can be expressed in Hamiltonian form: 
\be
   u_{\tau_m}=(-1)^{m}\frac{\D}{\D y}\left(\frac12 \frac{\delta C_{m+1}[u]}
                                                          {\delta u}
                                      \right).\label{um}
\ee
Here the functionals $C_m[u]$ represent the countable set of conservation laws
of KdV equation:
\be
   C_m=(-1)^m (2m+1)^{-1} \intl_{-\infty}^{\infty}dy\,
                          L^m\{y\,u_y(y,\tauvec)+2u(y,\tauvec)\},\label{Cm}
\ee
and the symbol $\delta/\delta u$ in (\ref{um}) denotes variational derivative:
$$
  \frac{\delta}{\delta u}=\suml_{j=0}^{\infty}\left(-\frac{d}{dx}\right)^j
                                               \frac{\D}{\D u_y^{(j)}},\qquad
  u^{(j)}_y\equiv\frac{\D^j u}{\D y^j}
$$

   In case of periodic function $u$ one should regard integrals in (\ref{Cm})
as integrals over period of $u$. But for us the most important fact is that
from (\ref{um}), (\ref{Cm}) it is readily seen that variational derivatives
of conservation laws $C_m[u]$ appear to be differential polynomials of
$u(y,\tauvec)$. Comparing (\ref{um1}) and (\ref{um}) we obtain
\be
   \veps_{t_m}=(-1)^m\frac{\delta C_m[u]}{\delta u}+\beta_m,\label{betam}
\ee
where $\beta_m$ are constants which in periodic case can't be deduced from
relations (\ref{um1}) and (\ref{um}). Computation of $\beta_m$ can be
carried out in each particular case (1-gap, 2-gap, etc.).

   Thus the meaning of the scheme of solutions construction modification
proposed in the present paper is in representation
of $f_m(y,\tauvec)=\veps_{t_m}$ in the form of differential polynomial
of $u(y,\tauvec)$ by virtue of (\ref{betam}). Then, knowing the solution
of KdV equation possessing desired spectral properties we can compute all
the necessary $f_m(y,\tauvec)$ and putting them in (\ref{general}) obtain
solution of the Harry Dym equation possessing the same spectrum.

In particular from (\ref{betam}) and (\ref{Cm}) it follows that
\ba
&& f_1(y,\tauvec)\equiv\veps_{t_1}=-2u+\beta_1 \label{f1}\\
&& f_2(y,\tauvec)\equiv\veps_{t_2}=6u^2-2u_{yy}+\beta_2 \label{f2}\\
\nonumber
&& f_3(y,\tauvec)\equiv\veps_{t_3}=-2[u^{(4)}_y-10uu^{(2)}_y-5(u_y)^2+10u^3]+\beta_3
\ea

\section{Construction of one-gap densities of polar operator}
\label{seconezone}

   In order to construct a family of one-gap densities of the polar operator
we start with smooth one-gap solution of the KdV equation of the form \cite{ZACH}
\be
   u(y,\tau_1)=-2\wp(i(y+4C\tau_1)+\omega)+{}^2/{}_3\,C.
   \label{uonezone}
\ee
Here $\wp(z)$ is elliptic Weierstrass function \cite{BE} defined in a standard
manner by real parameters $e_j,\ j=1,2,3$ such that $e_3<e_2<e_1$,
$\sum_j e_j=0$;
$\omega$, $\omega'$ are correspondingly real and purely imaginary
semi-periods of the function $\wp(z)$:
$\wp(\omega_{\alpha}|\,\omega,\omega')=e_{\alpha}$, $\alpha=1,2,3$, where
$$
  \omega_1=\omega,\quad \omega_2=\omega+\omega',\quad \omega_3=\omega',
$$
and $C\in\Real$ is arbitrary constant.

   Since in what follows we are interested in the densities of the polar operator
rather than solutions of the Harry-Dym equation, everywhere below we omit the
dynamics simply putting $\tauvec=0$. However it should be noted that
eqs. (\ref{general})
have been derived from HDE hierarchy and completely are based on dynamics.

   The spectrum of the Schr\"odinger equation (\ref{Schredinger})
(with potential
$u(y)=u(y,0)$) has only one gap $(E_1,E_2)$ and the spectrum boundary points
have form $E_0=e_3+{}^2/{}_3\,C$, $E_1=e_2+{}^2/{}_3\,C$,
$E_2=e_1+{}^2/{}_3\,C$.

  To obtain all 1-gap densities of the polar operator generated by the potential
(\ref{uonezone}) one has to insert (\ref{uonezone}) and (\ref{f1}) into
(\ref{general}). This yields
$$
  R(y)=\frac{C}{\wp(iy+\omega)+\frac23\,C+\frac{\beta_1}{4}}.
$$
Now one has to check the validity of relation
$u=\left(R^{-1/2}\right)_{yy}\,R^{1/2}$ where $u$ is given by (\ref{uonezone}).
It turns out that it holds not for any values of $C$ and $\beta_1$. Namely
we come to the following

\bf Lemma 1.\rm\  Potential
$u$ from (\ref{uonezone}) gives rise to density of the polar operator only if
$\hat{C}\equiv {}^2/{}_3\,C$ is solution of the equation:
$$
  4\hat{C}^3-\g_2\,\hat{C}+\g_3=0 
$$
and $\beta_1=0$.

Since the solution of the latter equation reads
$\hat{C}=-e_{\alpha}$, $\alpha=1,2,3$ one sees that there exist 1-gap
densities only with following arrangement of spectrum boundary points:
$E_0=e_3-e_{\alpha}$, $E_1=e_2-e_{\alpha}$, $E_2=e_1-e_{\alpha}$.
So by all means one of the boundary points coincides with $0$.

Expressions for finite-gap densities
(emerging if we put $t_1=0$ in solution of the Harry Dym equation) read:
\ba
\nonumber
&&   r_{\alpha}(x)=A_{\alpha}[e_{\alpha}-\wp(iy+\omega_1+\omega_{\alpha})],\\
&&   x=e_{\alpha}\,A_{\alpha}\,y-iA_{\alpha}[\zeta(iy+\omega+\omega_{\alpha})-\eta-\eta_{\alpha}],
   \quad \alpha=1,2,3.
   \label{Ronezone2}
\ea
Here
$$
  A_{\alpha}=\frac{{}^3/{}_2\,e_{\alpha}}{H_{\alpha}^2},\quad
  H_{\alpha}^2=(e_{\alpha}-e_{\beta})(e_{\alpha}-e_{\gamma}),\quad \alpha=1,2,3,
$$
$\zeta$ is Weierstrass $\zeta$-function: $-\zeta'(z)=\wp(z)$ and
$\eta_{\alpha}=\zeta(\omega_{\alpha})$.

   Let us briefly overview functional properties of constructed densities.

   The density $r_{\alpha}(x)$, $\alpha=1,2,3$ is periodic function with real
period
\be
   T_{\alpha}=e_{\alpha}\,A_{\alpha}2|\omega'|-2iA_{\alpha}\,\eta_3,
   \label{periodfin}
\ee
it's spectrum consists of spectral bands $[E_0,E_1]$ and $[E_2,+\infty]$.

   In case $\alpha=3$ one has: $E_0=0$, $E_1=e_2-e_3$, $E_2=e_1-e_3$.
Function $r_3(x)$ is even (i.~e. $r_3(-x)=r_3(x)$), periodic,
smooth and doesn't turn into $0$.
Therefore the string density arising in eq. (\ref{problem})
$\rho_3={}^1/{}_{r_3^2}$, possesses the same properties. It is density
which M.~G.~Krein wrote about in his paper \cite{KR} and therefore we refer to
function $r_3(x)$ as the Krein density. According to Krein's theorem there
exists just one periodic even density of the string equation
$\rho={}^1/{}_{r^2}$
having one-gap spectrum which starts at point $0$.
It should be noted that the density $r_3(x)$ was first constructed
in \cite{DM1}. It's period $T_3$ first had been found in \cite{BORD}
in terms of elliptic Jacobi functions.
It is this density which has
been constructed here explicitly.

   In case $\alpha=1$ $r_1(x)$ is a discontinuous function. Therefore
instead of smooth generating solution of the
KdV equation (\ref{uonezone}) we take singular one, namely
$$
  u(y,\tau_1)=-2\wp(i(y+4C\tau_1))+{}^2/{}_3\,C.
$$
Using the same technique we arrive to the density
\ba
\nonumber
&& \hat{r}_1(\hat{x})=A_1[e_1-\wp(iy+\omega)],\\
\nonumber
&& \hat{x}=e_1\,A_1\,y-iA_1[\zeta(iy+\omega)-\eta],
\ea

   As formula (\ref{periodfin}) for periods of densities $r_{\alpha}(x)$ remains
to be valid in this case we conclude that density $\hat{r}_1(\hat{x})$ is a periodic
even function with period $T_1=2e_1\,A_1\,|\omega'|-2iA_1\eta'$ having singularities
of the "cusp" type:
$$
  \hat{r}_1(\hat{x})\sim\mbox{Const}\,\hat{x}^{{}^2/{}_3}
$$
in points
$\hat{x}_n=nT_1$, $n\in\Integer$. Therefore we refer to this density as
cusp-periodic density. The boundary spectrum points are
$E_0=e_3-e_1$, $E_1=e_2-e_1$, $E_2=0$.

   This density is remarkable for it possesses degeneration property, namely
if the first spectral zone shrinks into single point ($E_0=E_1=-3\gamma$) and
continuous spectrum fills $\Real^{+}$ ($E_2=0$) then $\hat{r}_1(\hat{x})$
turns into well-known soliton density of the Harry Dym equation:
\ba
\nonumber
&& \hat{R}_1(y)=\th^2(\sqrt{3\gamma}\,y),\\
\nonumber
&& \hat{E}_1(y)=\frac{1}{\sqrt{3\gamma}}\,\th(\sqrt{3\gamma}\,y).
\ea

   In case $\alpha=2$ we obtain singular density similar to $r_1$.
Again changing the initial generating solution of the KdV equation by
$$
  u(y,\tau_1)=-2\wp(i(y+4C\tau_1)+\omega')+{}^2/{}_3\,C
$$
we arrive to the density
\ba
\nonumber
&& \hat{r}_2(\hat{x})=A_2[e_2-\wp(iy+\omega)],\\
\nonumber
&& \hat{x}=e_2\,A_2\,y-iA_2[\zeta(iy+\omega)-\eta].
\ea
Here the properties are quite similar to the properties of the $\hat{r}_1$.
Namely it has "cusps" at the points $\hat{x}=\frac{T_2}{2}+n\,T_2$,
where $T_2$ is it's period with respect to $\hat{x}$. Here the boundary
spectrum points are $E_0=e_3-e_2$, $E_1=0$, $E_2=e_1-e_2$.

   Note that when trying to degenerate $\hat{r}_2$ while the first spectral zone
shrinks into single point ($E_0=E_1=-3\gamma$) we get $A_2\to\infty$ and thus
$\hat{r}_2$ doesn't possess degeneracy property analogous to that of $\hat{r}_1$.

\section{Two-gap densities of the polar operator}\label{twozone}

   In this section we apply the technique described in sect. \ref{scheme} to
construct elliptic two-gap densities of the polar operator. As far as we know these
densities weren't present in literature before.

   Consider the case when the Riemann surface (\ref{curve}) possesses a special
sort of symmetry, namely let
\be
E_0=-\sqrt{3\g_2}+3a,\quad E_{\alpha}=-3e_{\alpha}+3a,\quad \alpha=1,2,3,\quad
E_4= \sqrt{3\g_2}+3a,                   \label{spectr}
\ee
where $a$ --- arbitrary real parameter.
The points of additional spectrum $\{\mu_k\}_{k=1}^2$
necessary to define 2-gap potential of the Schr\"odinger equation unambiguously
we choose (following \cite{EN}) to be the following:
$$
  \mu_1=E_2=-3e_2+3a,\quad \mu_2=E_3=-3e_3+3a,
$$
i.~e. we put additional spectrum at the ends of the "main" spectrum. Then
according to \cite{EN} the general formula for potential in terms
of 2-dimensional $\Theta$-functions is reduced to to form:
\be
   u(y)=-6\wp(iy+\omega|\,\omega,\omega')+3a,              \label{u}
\ee
i.~e. it is two-gap Lame potential.

In 2-gap case insertion of (\ref{u}), (\ref{f1}) and (\ref{f2}) into
(\ref{general}) yields
\be
   R(y)=\frac{K}{\wp^2(iy+\omega)+a\wp(iy+\omega)+b},   \label{R}
\ee
where
$$
b=a^2-\frac{\g_2}{4}+\frac{5}{24}\,a\beta_1+\beta_2,\quad
K=\frac{15}{18}\,a^2-\frac{7}{24}\,\g_2.        
$$

   Now as in 1-gap case one has to check the validity of the relation
$u(y)=\left(R^{-1/2}\right)_{yy}\,R^{1/2}$ with $u$ being the generated 2-gap
potential (\ref{u}). The following statement is valid:

\bf Lemma 2.\rm\  Potential
$u$ from (\ref{u}) gives rise to density of the polar operator only if
$a$ is solution of the equation:
$$
  \Big(a^2-\frac{\g_2}{3}\Big)(4a^3-\g_2 a-\g_3)=0,
$$
and the constants $\beta_1$ and $\beta_2$ are related as follows:
$\frac{5}{24}a\beta_1+\beta_2=0$.

One sees that parameter $a$ may take only 5 values:
$$
  a=\pm\sqrt{\frac{\g_2}{3}},e_1,e_2,e_3.
$$
This means that we get 5 two-gap densities of the form (\ref{R}).
Due to eqs. (\ref{spectr}) one of the spectrum boundary points by all means
coincides with zero.

   Each density $r_{\alpha}(x)$, $\alpha=\pm,1,2,3$ as in 1-gap cases
is defined by pair of equations
$$
  r_{\alpha}(x)=R_{\alpha}(y),\quad
  x=y-\intl_{0}^{y}(1-R_{\alpha}(y')\,dy'. 
$$

In the case $a^2=\pm\sqrt{\frac{\g_2}{3}}$ ($\alpha=\pm$) we obtain two densities
\ba
&&  r_{\pm}(x)=\frac{\g_2/3}{\left(\wp(iy+\omega)\pm\frac12\,\sqrt{\g_2/3}\right)^2},
\label{rpm1}\\
&&  x=y\,\frac{a^3}{a^3-\g_3}+
         \frac{ia^2}{a^3-\g_3}\{\zeta(iy+\omega-\gamma_{\pm})+\zeta(iy+\omega+\gamma_{\pm})-2\eta\}.
\label{rpm2}
\ea
where $\gamma_{\pm}$ is a solution of the equation
$\pm\frac{1}{2}\sqrt{\frac{\g_2}{3}}=-\wp(\gamma_{\pm})$.

The periods of functions $r_{\pm}(x)$ implicitly described by (\ref{rpm1}) and
(\ref{rpm2}) are
$$
  T_{\pm}=\frac{2a^2}{a^3-\g_3}(a|\omega'|+2i\eta'),\quad  a=\pm\sqrt{\frac{\g_2}{3}}.
$$

In the case $a=e_{\alpha}$ ($\alpha=1,2,3$) we obtain densities
\ba
&&  r_{\alpha}(x)=\frac{\frac{15}{8}e_{\alpha}-\frac{7}{24}\g_2}
                      {(\wp(iy+\omega)-e_{\beta})(\wp(iy+\omega)-e_{\gamma})},
                      \label{ralpha1}\\
\nonumber
&&  x=-y\cdot\left(A\left(\frac{e_{\beta}}{H_{\beta}^2}-
                               \frac{e_{\gamma}}{H_{\gamma}^2}\right)\right)-\\
\nonumber
        &&-iA\left(\frac{1}{H_{\gamma}^2}\zeta(iy+\omega+\omega_{\gamma})-
                   \frac{1}{H_{\beta}^2}\zeta(iy+\omega+\omega_{\beta})\right)
        +iA\left(\frac{\eta_{\gamma}}{H_{\gamma}^2}-\frac{\eta_{\beta}}{H_{\beta}^2}\right),\\
\nonumber
        &&A=\frac{K}{e_{\beta}-e_{\gamma}}.
\ea

Here it is assumed that symbols $\{\alpha,\beta,\gamma\}$ represent some transposition
of numbers $\{1,2,3\}$

   Each constructed density $r_{\alpha}(x)$, $\alpha=1,2,3$ is
a periodic function with period:
$$
   T_{\alpha}=2A|\omega'|\left(\frac{e_{\gamma}}{H_{\gamma}^2}-
                               \frac{e_{\beta}}{H_{\beta}^2}\right)-
              2Ai\eta'\left(\frac{1}{H_{\gamma}^2}-
                            \frac{1}{H_{\beta}^2}\right),\quad\alpha=1,2,3
$$

Among these five densities the most interesting is $r_{+}(x)$. It is smooth,
even, nonvanishing function. All of it's spectral bands $[E_0,E_1]$,
$[E_2,E_3]$, $[E_4,+\infty]$ lie on $\Real^+$ and $E_0=0$. Thus the density
$r_{+}(x)$ is two-gap density of Krein's type \cite{KR}.

Another four densities are discontinuous functions. Their  analysis is out of
scope of this paper.

\section{Auto-B\"acklund transformations of finite-gap densities}\label{BackSpec}

   In this section we analyze the densities which can be obtained by
auto-B\"acklund transformation of the one- and two-gap densities constructed
above.
Let's return for a while to a KdV and HD equations. Remind the following
statements \cite{RN}, \cite{DM1}. The pair of KdV solutions $u(y,t)$ and
$\hat{u}(y,t)$ related by the auto-B\"acklund transformation can be
written in terms of the HDE solution $r(x,t)=R(y,t)$ as
$$
  \hat{u}=u+\D_y^2\,\ln R,  
$$
where $R$ corresponds to $u$ under the relation
$u=\left(R^{-1/2}\right)_{yy}\,R^{1/2}$.

Let by $\hat{r}(\hat{x},t)=\hat{R}(y,t)$ denote the HDE solution related to
$\hat{u}$ in the same manner, namely:
\be
   \hat{u}=\left(\hat{R}^{-1/2}\right)_{yy}\,\hat{R}^{1/2}.\label{Rhat}
\ee
Then as follows from \cite{RN}, \cite{DM1}
\ba
 &&  \hat{R}=\frac{b}{R},\\ \label{Rhat2}
\nonumber
 &&  \hat{x}=y-\intl_0^y(1-\hat{R}(y',t))\,dy'.
\ea

In case of decreasing density ($r\to 1$ when $|x|\to\infty$) the constant $b=1$
\cite{RN}.
In case of periodic density this constant can not be obtained from the general
relations and some additional considerations are needed. It turns out that
constant $b$ values are different for individual densities. Below we construct
explicitly auto-B\"acklund transformation for 1- and 2-gap densities presented
above and compute the value of constant $b$ in each case.

\centerline{\underline{\bf 1-gap case}}

   Here we compute auto-B\"acklund transformation for smooth 1-gap
Schr\"odinger potentials, $u_{\alpha}(y)=-2\wp(iy+\omega)-e_{\alpha}$
which has been used in section \ref{seconezone} for generating the
densities $R_{\alpha}(y)=r_{\alpha}(x)$ given by (\ref{Ronezone2}).

From (\ref{Rhat}), (\ref{Rhat2}) it follows that
\be
\hat{u}_{\alpha}(y)=\left(R_{\alpha}^{{}^1/{}_2}(y)\right)_{yy}\,
                          R_{\alpha}^{-{}^1/{}_2}(y).\label{hatu}
\ee

The result of calculations is following. Potentials $\hat{u}_{\alpha}(y)$
are:
$$
  \hat{u}_{\alpha}(y)=-2\wp(iy+\omega+\omega_{\alpha})-e_{\alpha}.
$$

   When $\alpha=3$ we obtain smooth potential
which differs from its auto-B\"acklund transformation $u_3(y)$ by
shift in $y$.
When $\alpha=1,2$ we obtain singular potentials.
Thus in 1-gap case the auto-B\"acklund transformation for the KdV equation
doesn't lead out the class of known 1-gap potentials.

   In order to calculate auto-B\"acklund transformation
$\hat{r}_{\alpha}(\hat{x})=\hat{R}_{\alpha}(y)$
of 1-gap densities $r_{\alpha}(x)=R_{\alpha}(y)$ the whole procedure of densities
$R_{\alpha}$ construction should be repeated starting from potentials $\hat{u}_{\alpha}$
rather than $u_{\alpha}$.
The properties of densities $\hat{r}_1(\hat{x})$ and $\hat{r}_2(\hat{x})$
have been already considered in section \ref{seconezone}. The density
$\hat{r}_3(\hat{x})$
differs from the smooth density $r_3(x)$ which possesses no roots only by shift
in $x$ by ${}^{T_3}/{}_2$.

   Thus we have got 5 distinct 1-gap densities of the polar operator.
Densities $\hat{r}_1(\hat{x})$ and $\hat{r}_2(\hat{x})$ are continuous
cusp-periodic densities. Their auto-B\"acklund transformations $r_1(x)$ Й $r_2(x)$
correspondingly are singular functions. Densities $r_3(x)$ and $\hat{r}_3(\hat{x})$
differ only in shift by half-period ${}^{T_3}/{}_2$ in $x$ and represent
unique smooth 1-gap Krein density whose spectrum starts in $0$ and spectral zones
lie at $\Real^{+}$.

   Concluding analysis of auto-B\"acklund transformation in 1-gap case
we put expressions for the constant $b$ from (\ref{Rhat2}). Expressions are:
$$
  b\equiv b_{\alpha}=\frac{({}^3/{}_2\,e_{\alpha})^2}
                          {(e_{\alpha}-e_{\beta})(e_{\alpha}-e_{\gamma})},
  \quad \alpha=1,2,3.
$$

\centerline{\underline{\bf 2-gap case}}

   The scheme of calculating auto-B\"acklund transformation of two-gap
densities $r_{\alpha}(x)=R_{\alpha}(y)$, $\alpha=\pm,1,2,3$
(see eqs. (\ref{rpm1}), (\ref{ralpha1})) is the following:
one computes $\hat{u}_{\alpha}(y)$ by means of relation (\ref{hatu})
and then use this function as generating Schr\"odinger potential
in the scheme of constructing polar operator densities presented
in section \ref{seconezone}.
The result reads:
\ba
\nonumber
&& \hat{r}_{\pm}(\hat{x})=K_{\pm}\Big(\wp(iy+\omega)\pm\frac12\,\sqrt{\g_2/3}\Big)^2\\
\nonumber
&& \hat{x}=\frac{K_{\pm}\,\g_2}{6}\,y-\frac{iK_{\pm}}{6}\,\wp'(iy+\omega)\mp iaK_{\pm}[\zeta(iy+\omega)-\eta]
\ea
where
$$
  K_{\pm}=\frac{75a^2-7\g_2}{12a(a^3-\g_3)a^2},\quad a=\pm\sqrt{\frac{\g_2}{3}}.
$$

Another three densities have the form:
\ba
\nonumber
&&\hat{r}_{\alpha}(\hat{x})=K_a(\wp(iy+\omega)-e_{\beta})(\wp(iy+\omega)-e_{\gamma}),
  \quad \alpha=1,2,3.\\
\nonumber
&& \hat{x}=K_{\pm}\,\Big(\frac{\g_2}{12}+e_{\beta}\,e_{\gamma}\Big)\,y-
           \frac{iK_{\pm}}{6}\,\wp'(iy+\omega)-ie_{\alpha}K_{\pm}[\zeta(iy+\omega)-\eta]
\ea
where
$$
  K_{\alpha}=\frac{(75a^2-7\g_2)}{2(3a^2-\g_2)(12a^2-\g_2)
                            \left(\frac{15}{8}e_{\alpha}-\frac{7}{24}\g_2\right)},\quad
  a=e_{\alpha},\alpha=1,2,3
$$

Before describing properties of the densities obtained let us note the
following. Since we don't prove the fact that auto-B\"acklund transformation
in periodic case preserves the structure of spectrum of the corresponding
linear problem for the polar operator we can not state that spectra of the
densities $\hat{r}_{\pm}(\hat{x})$, $\hat{r}_{\alpha}(\hat{x})$ consist
of only two bands. Nevertheless properties of these densities outlined
below suggest the validity of the said hypothesis.

Now we briefly describe properties of the obtained densities.
The density $\hat{r}_+(\hat{x})$ is the smooth periodic symmetric nonvanishing
density of the Krein type. The density $\hat{r}_-(\hat{x})$ has cusp-type
singularities
$$
\hat{r}_-(\hat{x})
\mathbin{\raisebox{-1pt}{${}_{\widetilde{\hat{x}\to \hat{x}_0}}$}}
\mbox{Const}\,(\hat{x}-\hat{x}_0)^{{}^4/{}_5}.
$$
The periodic densities $\hat{r}_{\alpha}(\hat{x})$ also have cusp singularities although
of another type:
$$
\hat{r}_{\alpha}(\hat{x})
\mathbin{\raisebox{-1pt}{${}_{\widetilde{\hat{x}\to \hat{x}_{\alpha}}}$}}
\mbox{Const}\,(\hat{x}-\hat{x}_{\alpha})^{{}^2/{}_3}.
$$

Let us note that to implement the whole scheme which led to the above
expressions the constants $\beta_1$ and $\beta_2$ in (\ref{f1}) and (\ref{f2})
should satisfy condition
$$
  (30a\beta_1+\beta_2)=180a^2.
$$

Finally calculations for the constants $b$ in (\ref{Rhat2}) for each pair of
densities linked by the auto-B\"acklund transformation give:
\ba
\nonumber
&& b_{\pm}=\frac{75a^2-7\g_2}{12a(a^3-\g_3)a^2},\quad
   a=\pm\sqrt{\frac{\g_2}{3}}\\
\nonumber
&& b_{\alpha}=\frac{(75a^2-7\g_2)}{2(3a^2-\g_2)(12a^2-\g_2)
                            \left(\frac{15}{8}e_{\alpha}-\frac{7}{24}\g_2\right)},\quad
   a=e_{\alpha},\alpha=1,2,3.
\ea

\section{Acknowledgements}

The authors acknowledge RFBR Grant
\# 99-01-00696 for support of this work.


\end{document}